\documentclass{article}
\usepackage{spconf,amsmath,epsfig}

\usepackage[]{graphicx}
\usepackage{caption}
\usepackage{subfig}
\usepackage{amsmath}
\usepackage{amssymb}
\usepackage{epsfig}
\usepackage{cite}
\usepackage{color}
\usepackage{balance}
\usepackage{amsmath}
\usepackage{amsfonts} 
\usepackage{graphicx}
\usepackage{pdfpages}
\usepackage[T1]{fontenc} 
\usepackage{array}
\usepackage{url}

\usepackage{color}
\usepackage{wrapfig}
\usepackage{lipsum}
\usepackage[hidelinks]{hyperref}
\usepackage{multirow}
\usepackage{tabularx}
\usepackage{gensymb} 
\usepackage{breqn}
\usepackage{array}
\usepackage{booktabs}

\title{A Potential Low Cost remote sensing using GPS derived PWV}

\name{Shilpa Manandhar$^{1}$, Yee Hui Lee$^{1}$, Yu Song Meng$^{2}$, Feng Yuan$^{1}$ and  Soumyabrata Dev$^{3}$
\thanks{This research is funded by the Defence Science and Technology Agency (DSTA), Singapore.}
\thanks{Send correspondence to Y.\ H.\ Lee, E-mail: EYHLEE@ntu.edu.sg.}
}
\address{
	$^{1}$~School of Electrical and Electronic Engineering, Nanyang Technological University, Singapore \\
	$^{2}$~National Metrology Centre, Agency for Science, Technology and Research (A$^{*}$STAR), Singapore \\     
    $^{3}$~The ADAPT Centre, School of Computer Science and Statistics, Trinity College Dublin, Ireland \\
}

\begin{document}

\maketitle 
\begin{abstract}
In this paper, the Precipitable Water Vapor (PWV) content of the atmosphere is derived using the Global Positioning System (GPS) signal delays. The PWV values from GPS are calculated at different elevation cut-off angles. It was found that the significant range of elevation cut-off angles is from 5$^{\circ}$ to 50$^{\circ}$. The PWV values calculated from GPS using varying cut-off angles from this range were then compared to the PWV values calculated using the radiosonde data. The correlation coefficient and the Root Mean Square (RMS) error between the GPS and radiosonde derived PWV decreases with the increasing cut-off angle and the distance between the two. The seasonal parameters also effect the relation between the two.
\end{abstract}

\begin{keywords}
PWV, Remote Sensing, GPS, GPS Elevation Cut-Off Angles.
\end{keywords}

\section{Introduction}
\parskip -1pt 
\label{sec:intro}
Precipitable Water Vapor (PWV) is the depth of water in a column of the atmosphere, if all the water in that column were precipitated as rain. PWV is generally expressed in terms of mm. PWV values can be derived using radiosondes, microwave radiometers, very long baseline interferometry (VLBI), satellite based instruments and Global Positioning System (GPS) signal delays. The water vapor climatology is typically investigated using radiosonde and satellite observations. But such observations have limitations in capturing varied data with good resolution as they have low spatial and temporal resolutions~\cite{RadiosondeDisadv}. 

Ground-based GPS meteorology, on the other hand, offers improved spatial and temporal resolutions for studying the water vapor variations compared to the traditional techniques. Thus, with the rapid deployment of GPS monitoring stations in local, regional, and global scales, GPS is extensively being used as an all-weather, low-cost remote sensing instrument and the PWV values derived from GPS signals are useful in weather forecasting and climatology \cite{IGARSS_2016}.

The accuracy of PWV values derived from GPS (\textit{GPS-PWV}) can be verified by comparing it to that obtained from other diverse sources [3]. The general practice found in literature is to compare the \textit{GPS-PWV} with the PWV values calculated using radiosonde data (\textit{Rad-PWV}) recorded by the collocated radiosonde station. The \textit{GPS-PWV} and \textit{Rad-PWV} have been shown to have a good correlation when the stations are collocated [4]. Here it is interesting to note that the \textit{GPS-PWV} represents the PWV value which is an average over a cone. The cone size depends on the different elevation angles for GPS observations, whereas the \textit{Rad-PWV} value is the integration over a single path of the radiosonde balloon. Thus, in this paper~\footnote{The source code of all simulations in this paper is available online at \url{https://github.com/shilpa-manandhar/GPS-cutoff-angles}}, we present the correlation between \textit{Rad-PWV} and \textit{GPS-PWV} at varying cut-off elevation angles of GPS observations. 

In the following sections, the locations of different GPS and radiosonde stations are described in brief. Subsequently, we discuss the derivation of \textit{GPS-PWV} and \textit{Rad-PWV}. This is followed by the detailed comparison between the \textit{GPS-PWV} and \textit{Rad-PWV}. Finally, we conclude the paper with the important findings and share the future perspectives.

\section{Station Locations and Databases}
One year (year 2015) of data from 1 Radiosonde station and 4 GPS stations around Singapore have been used in this paper. The GPS and the radiosonde stations are indicated on the map in Fig.~1 with corresponding GPS station names. The radiosonde data for the given station is downloaded form the database of Wyoming University, which provides radiosonde data for any chosen station twice a day [5]. The GPS files for these four stations have been provided by Singapore Satellite Positioning Reference Network (SiReNT). The temporal resolution of \textit{GPS-PWV} is 5 min. 

\begin{figure}[t!]
\begin{center}
\includegraphics[width=0.42\textwidth]{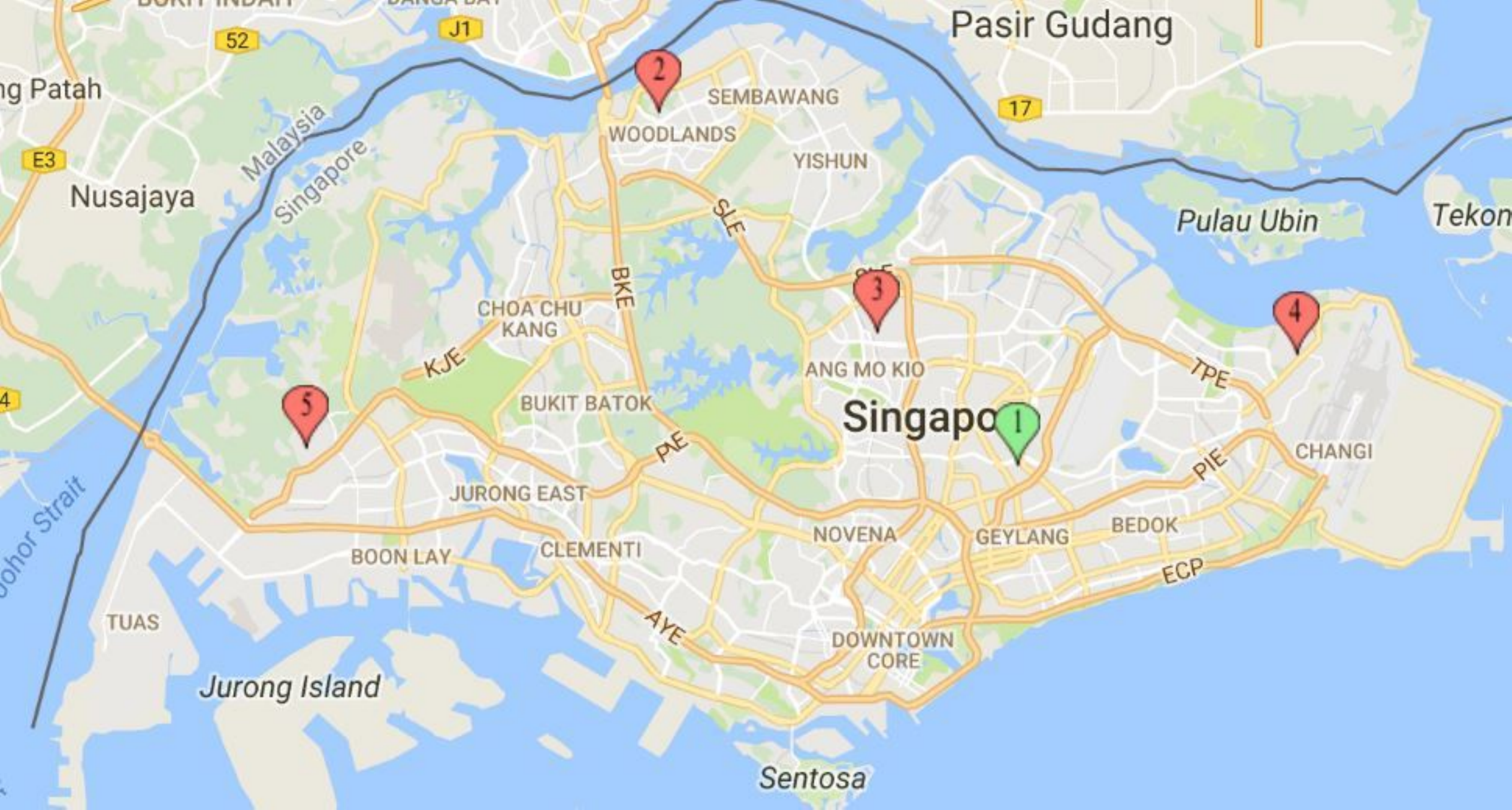}
\caption{Map of Singapore showing 4 GPS Stations (red marker) and 1 Radiosonde Station (green marker). Marker 1: Radiosonde Station, GPS Station Names; Marker 2: SRPT, Marker 3: SNPT, Marker 4: SLYG and Marker 5: SNYU.
\label{fig:map}}
\end{center}
\vspace{-0.8cm}
\end{figure} 

\section{Derivation of PWV Values}
In this section, the methods to derive PWV values from GPS signal delay and radiosonde data have been discussed. 

\subsection{PWV Derived from GPS}
The PWV values (in mm) are calculated using the zenith wet delay (ZWD), \textit{$\delta$L$_w^{o}$}, incurred by the GPS signals as shown in eq.\ref{eq1}. 

\begin{equation}
	\mbox{PWV}=\frac{PI \cdot \delta L_w^{o}}{\rho_l}
    \label{eq1}
\end{equation}     

\begin{dmath}
	PI=[-1\cdot sgn(L_{a})\cdot 1.7\cdot 10^{-5} |L_{a}|^{h_{fac}}-0.0001]\cdot cos(\frac{(DoY-28)2\pi}{365.25})+[0.165-(1.7\cdot 10^{-5})|L_{a}|^{1.65}]+f
    \label{eq2}
\end{dmath}

Here, $\rho_{l}$ is the density of liquid water (1000 kg$/m^{3}$). \textit{PI} is the dimensionless factor determined by eq.~\ref{eq2}  \cite{shilpaPI} , where, \textit{L$_{a}$} is the latitude,  \textit{DoY} is day-of-year, \textit{h$_{fac}$} is 1.48 for stations from northern hemisphere and 1.25 for stations from southern hemisphere.  $f=-2.38\cdot 10^{-6}H$, where \textit{H} is the station height and factor  \textit{f} can be ignored for stations with height less than 1000 m. 
\begin{figure}[htb]
\begin{center}
\includegraphics[width=0.3\textwidth]{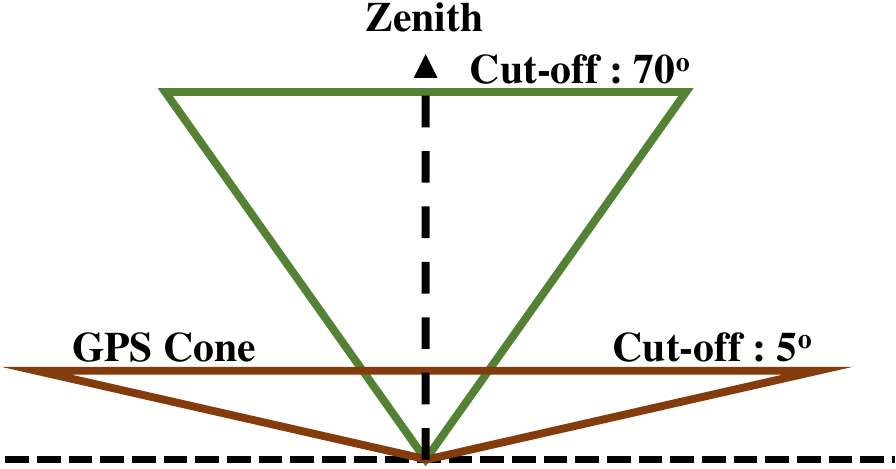}
\caption{Schematic diagram of elevation cut-off angles for GPS satellite observations.
\label{fig:gps-cone}}
\end{center}
\vspace{-0.5cm}
\end{figure}

The ZWD values are processed using the GIPSY OASIS software and recommended scripts \cite{GIPSY}. In the GIPSY OASIS software, the elevation cut-off angle for the GPS observation can be defined by the user. As shown in Fig. \ref{fig:gps-cone} smaller the cut-off angle, larger is the GPS cone. In this paper, the PWV values are calculated for all the 4 GPS stations using the station specific information in eq.(\ref{eq1}-\ref{eq2}) for a range of varying elevation cut-off angles.  

\subsection{PWV Derived from Radiosonde Data}
Water vapor pressure can be calculated using the radiosonde
temperature and relative humidity data as shown in eq. \ref{eq3}.

\begin{dmath}
e = RH \cdot exp(-37.2465+0.213166\cdot T-2.56908\cdot 10^{-4} \cdot T^{2}),
    \label{eq3}
\end{dmath}
where \textit{RH} is relative humidity in percentage, \textit{T} is the absolute temperature in Kelvin and \textit{e} is the water vapor pressure in hPa. By applying the gas state equation, the water vapor density, $\rho{_{v}}$ can be computed as given in eq.~\ref{eq4}.

\begin{dmath}
	\rho_{v}=\frac{e}{R_{v} \cdot T}
    \label{eq4}
\end{dmath}

Here, \textit{R}$_{v}$=461.525\textit{K}$^{-1}$\textit{kg}$^{-1}$ is the specific gas constant for water vapor. Now, the PWV values can be calculated as shown below,
\begin{dmath}
	PWV=(1/\rho_{l})\cdot \sum \frac{(h_{j+1}-h_{j}) (\rho_{v}^{j+1}-\rho_{v}^{j})}{2},
    \label{eq5}
\end{dmath}
where \textit{h$_{j}$} is the altitude at \textit{j$^{th}$} level \cite{RadPWV}. For this paper, the PWV values are calculated from the radiosonde data using eqs.~(\ref{eq3}-\ref{eq5}).

\section{Results and discussions}
\begin{figure}[htb]
\begin{center}
\includegraphics[width=0.35\textwidth]{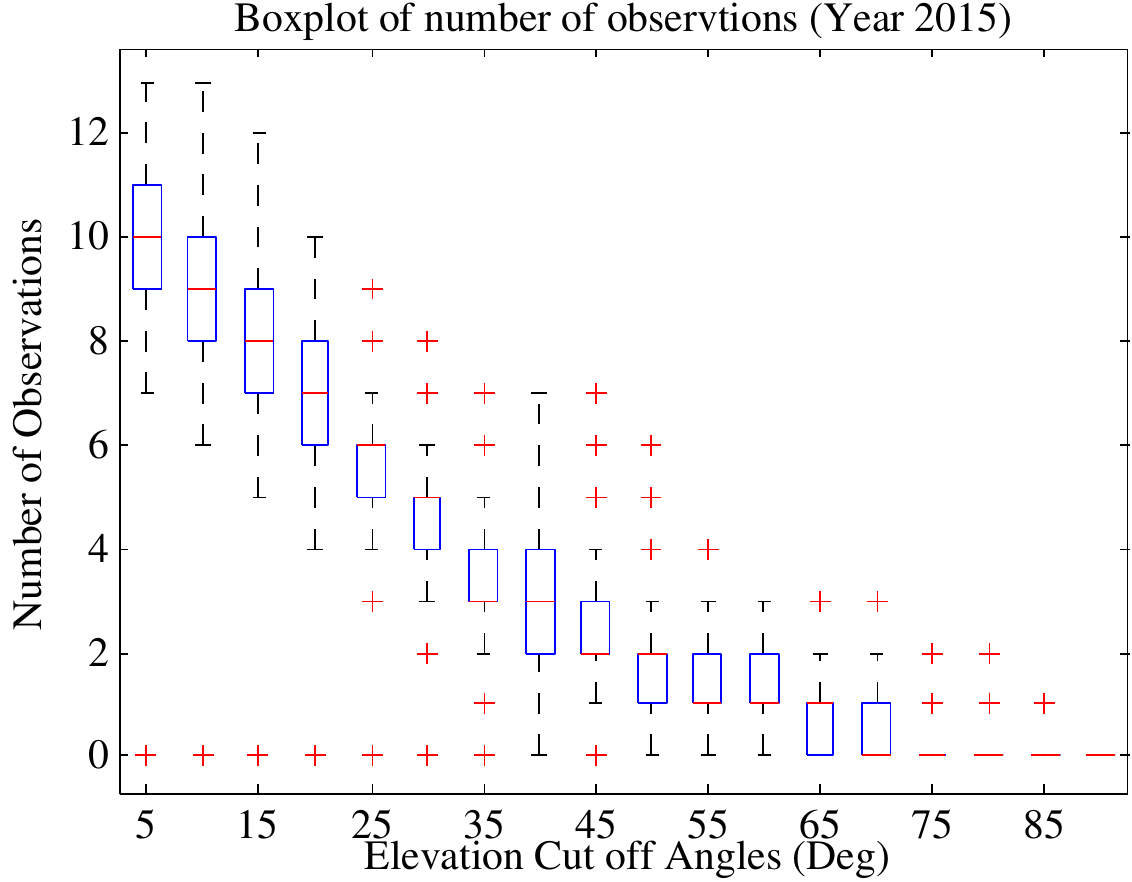}
\caption{Number of GPS observations at different Elevation Cut-Off Angles for both AM and PM timings of 2015 for SLYG (GPS station).
\label{fig:cut-off}}
\end{center}
\vspace{-0.7cm}
\end{figure} 
\vspace{-0.5cm}
\subsection{Methodology and Evaluation Metrics}
The \textit{GPS-PWV} values are calculated for the range of elevation angles for all the GPS stations as marked in Fig. 1. The \textit{GPS-PWV}  values are computed with the temporal resolution of 5 min. Similarly, the \textit{Rad-PWV} values are calculated for the single radiosonde station. The radiosonde data for Singapore is available twice a day; 00 UTC, referred to as AM, and 10 UTC, referred to as PM, in this paper. The \textit{GPS-PWV} values are then extracted at 00 and 10 UTC to match the radiosonde timings for the comparison purpose.

Correlation coefficient and Root Mean Square (RMS) error are two evaluation metrics that have been used to compare the \textit{GPS-PWV} and \textit{Rad-PWV} and analyze the effect of different range of angles and other factors.

\begin{figure*}[htb]
\begin{center}
\includegraphics[width=0.35\textwidth]{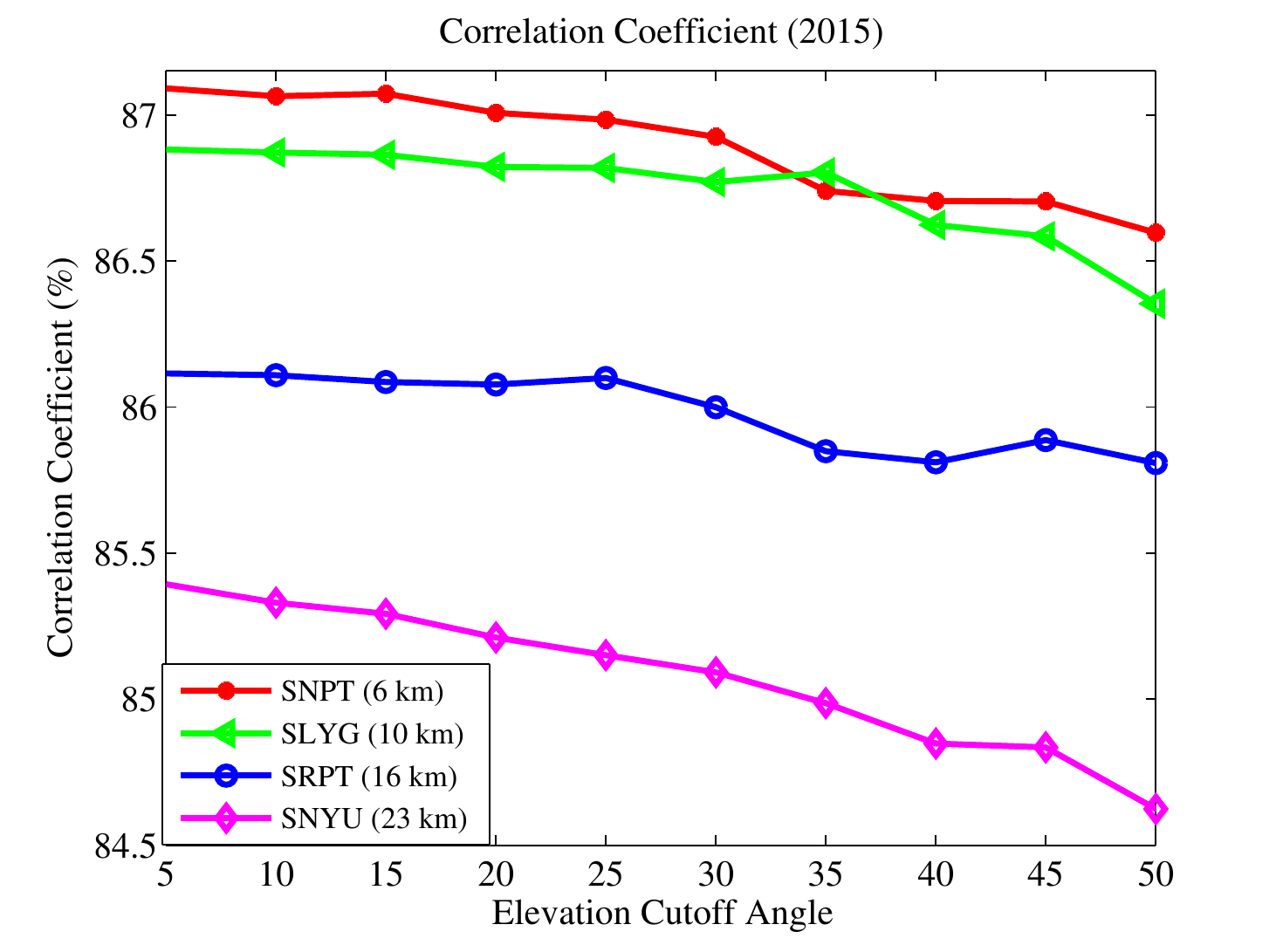}
\includegraphics[width=0.35\textwidth]{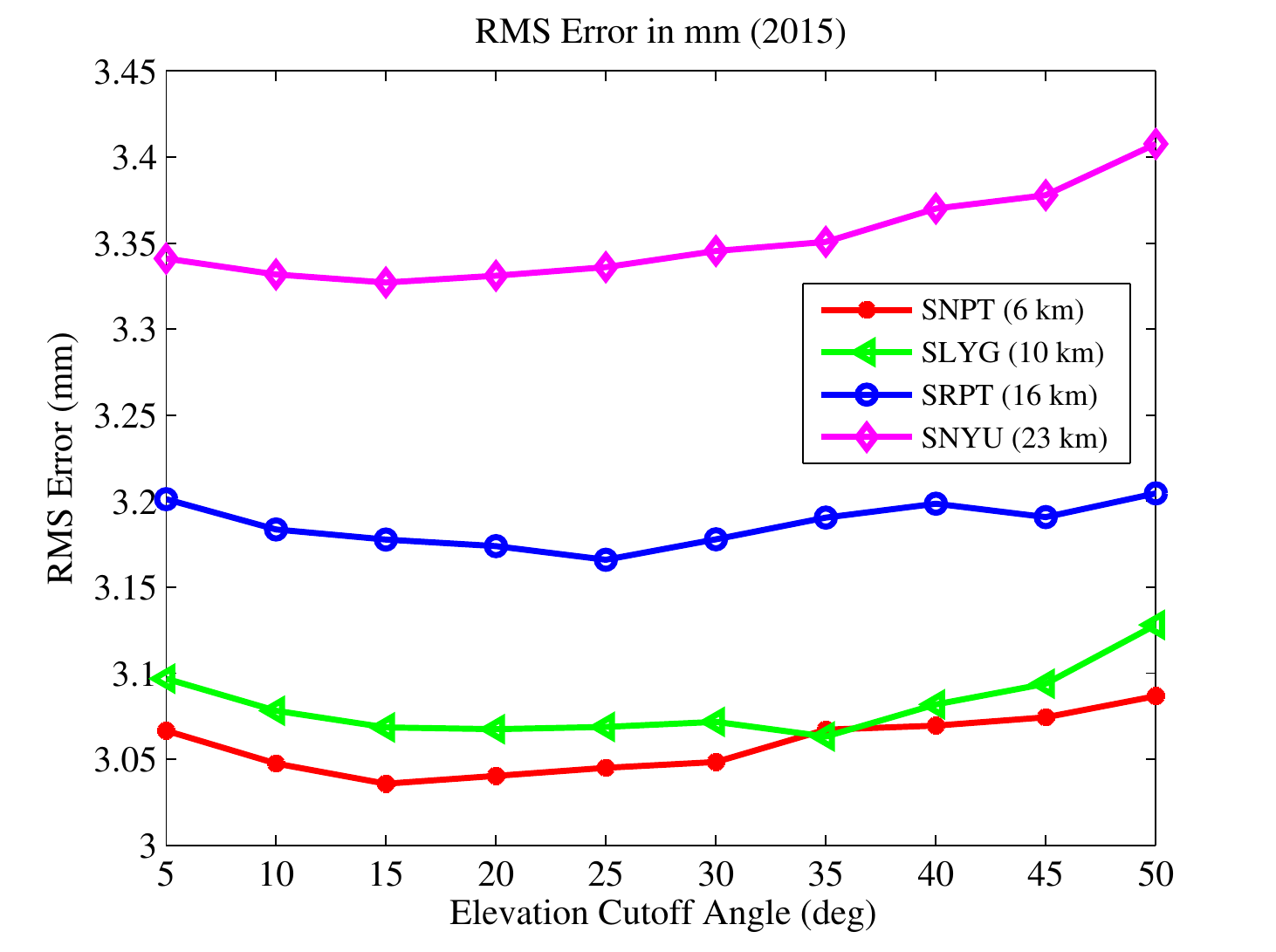}
\caption{Correlation Coefficient in \% (left) and RMS error in mm (Right) between \textit{GPS-PWV} and \textit{Rad-PWV}.
\label{fig:rms_error}}
\end{center}
\vspace{-0.9cm}
\end{figure*}

\vspace{-0.3 cm}
\subsection{Range of Elevation Cut-Off Angles for GPS PWV}
This section discusses about the range of elevation cut-off angles that has been used in this paper. As discussed before, \textit{GPS-PWV} is the volumetric average of multiple slant path PWV values. The number of slant paths depends on the number of satellite observations present inside the GPS cone. The GPS cone is defined by the chosen elevation cut-off angle. As the cut-off angle increases, the cone becomes narrower and the number of GPS satellite observations inside the cone becomes smaller (cf. Fig. \ref{fig:gps-cone}). There is a common practice of taking 7$^{\circ}$ to 10$^{\circ}$ of elevation cut-off angle for PWV calculations \cite{PWV_GPS_verify,Junbo}. 

Fig.~\ref{fig:cut-off} shows the number of GPS observations available when a particular elevation cut-off angle is chosen. The x-axis of the figure is the elevation cut-off angle and y-axis is the number of observations available for every 5 min reading within the particular cut-off angle. The figure shows the total number of observations for both the AM and PM timings from year 2015 for the GPS station, SLYG (cf. Fig. \ref{fig:map}). The boxplot clearly shows the range of the number of observations that is available for a particular cut-off angle. It can be seen that the number of observation decreases with the increase in the cutoff angle. That means the \textit{GPS-PWV} at lower cut-off angle represents the average of many number of slant paths, at maximum reaching up to 12 paths. At higher cut-off angle the \textit{GPS-PWV} may represent only a single slant path PWV value. At cut-off angle higher than 70$^{\circ}$, there is almost no observations. Thus in the following analysis, a cut-off range of 5$^{\circ}$ to 50$^{\circ}$ is presented as the PWV values calculated within this range represent the average of a substantial (at least 2 paths per reading) number of slant path PWV values. 

\vspace{-0.3 cm}
\subsection{GPS PWV Values at Different Cut-Off Angles}
The \textit{GPS-PWV} calculated at different elevation cut-off angles from 5$^{\circ}$ to 50$^{\circ}$ for the 4 GPS stations are compared against the \textit{Rad-PWV} values. The result is discussed in terms of the correlation coefficient (\%) and RMS error (mm). Fig. \ref{fig:rms_error} shows the average correlation coefficient and average RMS error for both the AM and PM values over a period of one year (2015). The legend in the figure
indicates the distance between a particular GPS station and the radiosonde station. The smallest distance is 6 km and the largest is 23 km. 

\vspace{-0.4cm}
\subsubsection{Trend in Correlation Coefficient}
From the correlation coefficient plot in Fig. \ref{fig:rms_error}, it can be observed that for all the GPS stations, the average correlation coefficient between \textit{Rad-PWV} and \textit{GPS-PWV} decreases with the increasing elevation cut of angles. The effect of the cut off angles is more prominent for stations far from the radiosonde station. The correlation coefficients for GPS stations namely SNPT, SLYG and SRPT (cf. Fig.~\ref{fig:map}) are quite similar. This is because the radiosonde balloon has been observed to have a drift of up to 15 km away from the radiosonde station. Therefore stations within 15 km of the radiosonde station are expected to have a similar trend. The 3 stations SNPT, SLYG and SRPT are at a distance of 6, 10 and 16 km respectively, and so similar trends are observed. The only GPS station SNYU, which is more than 15 km away from the radiosonde station (23 km), gives the smallest correlation coefficient. In general, as distance increases, correlation coefficient decreases as shown in Fig~\ref{fig:rms_error}.    

Here it is interesting to discuss the results from the literature \cite{EleCutOff}, which compares the \textit{GPS-PWV} at different elevation cut-off angles to the \textit{Rad-PWV} using data from Sweden. The results show that the correlation coefficient varies by a significant amount when the cut-off angle changes. One of the reasons is the distance between the radiosonde station and the GPS station is mostly greater than 50 km in \cite{EleCutOff}. Another important reason is that, unlike the stations in \cite{EleCutOff}, the GPS stations used in this paper lack the variation in altitude levels and Singapore has the tropical climate, thus not much variation in water vapor content is observed at different stations for the same period of time.
\vspace{-0.3 cm}

\subsubsection{Trend in RMS Error} 
The RMS error between the \textit{GPS-PWV} and \textit{Rad-PWV} for different stations at varying elevation cut-off angles are shown in the RMS error plot in Fig. \ref{fig:rms_error}. The trend in RMS error is similar to the correlation coefficient and can be described in similar manner. The RMS error for GPS stations nearer to the radiosonde station is smaller compared to the farther ones. It can be observed that the RMS error in the lowest elevation angle i.e. 5$^{\circ}$ is higher. This is because of the effect of multipath, which is high in the lower elevation cut-off angles as reported in \cite{EleCutOff}.
\vspace{-0.3 cm}
\subsubsection{Effect of Seasonal Parameters}
In section 4.3, the effect of cut-off elevation angles on the RMS error and correlation coefficient between \textit{Rad-PWV} and \textit{GPS-PWV} was shown. In this section, the effect of different seasonal and weather parameters on the evaluation metrics is discussed. The correlation coefficients (in \%) between \textit{Rad-PWV} and \textit{GPS-PWV} at a cut-off angle of 20$^{\circ}$ is listed in Table~\ref{table:corr}. 

It should be noted that a significant change is observed when the coefficients from AM and PM are compared. The correlation coefficient for PM are generally smaller than the AM for a given elevation angle. This is because, Singapore generally has Monsoon season with many rainfall events occurring in the evening. These evening rainfall events are mostly the convective rain events which occur for small amount of time, covering smaller area~\cite{JunXiang}. Hence, climate during the evening time is more localized in Singapore and is reflected upon the correlation coefficients.

\begin{table}[htb]
\centering
\caption{Correlation Coefficients (\%) at different times of day; AM and PM and at cut-off angle of 20$^\circ$ }
\small  
\begin{tabular}{*{6}{c}}
\toprule
\multicolumn{1}{c|}{Time} & \multicolumn{1}{c|}{\shortstack{SNPT \\ 6 km}} & \multicolumn{1}{c|}{\shortstack{SLYG \\ 10 km}}& \multicolumn{1}{c|}{\shortstack{SRPT \\ 16 km}}& \multicolumn{1}{c}{\shortstack{SNYU \\ 23 km}}\\
\midrule
{AM} & 95.6    & 95.7  & 95.6    & 94.3 \\
{PM} & 78.3    & 77.8  & 76.4    & 76.1 \\
\hline
\end{tabular}
\label{table:corr}
\vspace{-0.5 cm}
\end{table}

\section{Conclusion \& Future Work}
\label{sec:conc}
In this paper, \textit{GPS-PWV} values from 4 different stations have been calculated at varying elevation cut-off angles. The results shows that a substantial number of GPS observations can be found up to a cut-off angle of 50$^{\circ}$. Thus, the \textit{GPS-PWV} calculated by varying the cut-off angles from 5$^{\circ}$ to 50$^{\circ}$ are used for comparison to the \textit{Rad-PWV} calculated from a single Radiosonde station. It was observed that for a small tropical island like Singapore, different cut-off angles have very small impact on the correlation coefficient and the RMS errors. At the same time, it was noticed that the correlation coefficients and the RMS errors have significant differences when the AM and PM observations are inter compared. This is due to the many convective rain events that is experienced mostly in the evening time during the monsoon seasons. 

In our future work, we plan to include data from various stations across the tropical regions. This will provide us a detailed statistical analysis of GPS derived PWV, helping in understanding the hydrological balance in nature. We will also use imaging data captured from the ground~\cite{WAHRSIS}, to further understand precipitation~\cite{rainonset} in atmosphere. 
 
\balance

\end{document}